\documentclass{pnastwo}

\usepackage{amssymb,amsfonts,amsmath}
\usepackage{xspace}
\usepackage[T1]{fontenc}
\usepackage{deluxetable}
\usepackage{graphicx}

\newcommand{\mse}{\mbox{m\,s$^{-1}$}}

\newcommand{\rearth}{\ensuremath{ R_{\oplus} }\xspace} 
 
\newcommand{\rearthe}{\ensuremath{ R_{\oplus} }\xspace} 
 
\newcommand{\mearth}{\ensuremath{ M_{\oplus} }\xspace} 
 
\newcommand{\mearthe}{\ensuremath{ M_{\oplus} }\xspace}

\newcommand{\ek}{\textit{Kepler\ }}

\newcommand{\gcc}{\mbox{g\,cm$^{-3}$}}

\newcommand{\araa}{ARA\&A}
\newcommand{\apj}{ApJ}
\newcommand{\apjl}{ApJLett}
\newcommand{\apjs}{ApJS}
\newcommand{\aap}{A\&A}
\newcommand{\mnras}{MNRAS}
\newcommand{\pasp}{PASP}
\newcommand{\nat}{Nature}
\contributor{Submitted to Proceedings
of the National Academy of Sciences of the United States of America}
\issuedate{Issue Date}
\volume{Volume}
\issuenumber{Issue Number}

\begin{document}



{\title{Occurrence and core-envelope structure of \\ 1--4x Earth-size
    planets around Sun-like stars}}

\author{Geoffrey W. Marcy\affil{1}{University of California, Berkeley},
Lauren M. Weiss\affil{1}{University of California, Berkeley},
Erik A. Petigura\affil{1}{University of California, Berkeley},
Howard Isaacson\affil{1}{University of California, Berkeley},
Andrew W. Howard\affil{2}{Institute for Astronomy, University of Hawaii}\and
Lars A. Buchhave \affil{3}{Harvard-Smithsonian Center for Astrophysics,}
}

\contributor{Submitted to Proceedings of the National Academy of Sciences
of the United States of America}

\maketitle

\begin{article}

\begin{abstract} 
  {Small planets, 1--4x the size of Earth, are extremely common around
    Sun-like stars, and surprisingly so, as they are missing in our
    solar system. Recent detections have yielded enough information
    about this class of exoplanets to begin characterizing their
    occurrence rates, orbits, masses, densities, and internal
    structures.  The {\em Kepler} mission finds the smallest planets
    to be most common, as 26\% of Sun-like stars have small, 1-2
    \rearth planets with orbital periods under 100 days, and 11\% have
    1--2 \rearth planets that receive 1-4x the incident stellar flux
    that warms our Earth.  These Earth-size planets are sprinkled
    uniformly with orbital distance (logarithmically) out to 0.4 AU,
    and probably beyond. Mass measurements for 33 transiting planets
    of 1--4 \rearth show that the smallest of them, $R < 1.5$
    \rearthe, have the density expected for rocky planets.  Their {\it
      densities increase with increasing radius}, likely caused by
    gravitational compression.  Including solar system planets yields
    a relation: $\rho = 2.32 + 3.19 R/R_{\oplus}$ [\gcc]. Larger
    planets, in the radius range 1.5--4.0 \rearthe, have {\it
      densities that decline} with increasing radius, revealing
    increasing amounts of low-density material (H and He or ices) in
    an envelope surrounding a rocky core, befitting the appellation
    ``mini-Neptunes.''  Planets of $\sim$ 1.5 \rearth have the highest
    densities, averaging near 10 \gcc.  The gas giant planets occur
    preferentially around stars that are rich in heavy elements, while
    rocky planets occur around stars having a range of heavy element
    abundances.  One explanation is that the fast formation of rocky
    cores in protoplanetary disks enriched in heavy elements permits
    the gravitational accumulation of gas before it vanishes, forming
    giant planets.  But models of the formation of 1--4 \rearth
    planets remain uncertain. Defining habitable zones remains
    difficult, without benefit of either detections of life elsewhere
    or an understanding of life's biochemical origins.}
 \end{abstract}


\section{Significance Statement}
Among the nearly 4000 planets known around other stars, the most
common are 1--4x the size of Earth.  A quarter of Sun-like stars have
such planets orbiting within half an Earth's orbital distance of them,
and more surely orbit farther out. Measurements of density show that
the smallest planets are mostly rocky while the bigger ones have rocky
cores fluffed out with hydrogen and helium gas, and likely water,
befitting the term ``mini-Neptunes.'' The division between these two
regimes is near 1.5 \rearthe. Considering exoplanet hospitality, 11\%
of Sun-like stars have a planet of 1--2x the size of Earth that
receives between 1.0--4.0x the incident stellar light that our Earth
enjoys.  However, we remain ignorant of the origins of, and existence
of, exobiology, leaving the location of the habitable zone uncertain.
\bigskip



\dropcap{N}ASA's \ek mission astonishingly revealed a preponderance of
planets having sizes between 1 and 4 times the diameter of Earth
\cite{Borucki2011, Batalha2013, Howard2013a, Petigura2013a,
  Fressin2013}. Our Solar System has no planets larger than Earth and
smaller than Neptune (3.9 \rearthe).
As such, these new planets are poorly understood. Uranus and Neptune
provide clues:\ they have rocky cores of $\sim$10 \mearth, enveloped
by a modest amounts of H and He gas. But the clues are limited by
the difficulty in explaining only modest amounts of gas with standard
models of runaway gas accretion in the protoplanetary disk
\cite{Pollack1996, Goldreich2004, Rogers_Seager2010b, Morbidelli2013}.
Planet formation models face another challenge as they predicted very
few planets with final sizes 1--4 \rearth \cite{Ida_Lin2010,
  Mordasini2012a, Alibert2013}.

This great population of sub-Neptune-mass exoplanets had first been
revealed by precise Doppler surveys of stars within 50 pc
\cite{Howard2010b, Mayor2011}, a finding that \ek's discoveries
confirm. While most of the over 3000 1--4 \rearth planets found by \ek
are officially only ``candidates,'' 90\% of those candidates are real
planets \cite{Morton_Johnson2011, Fressin2013, Marcy2014}. After
accounting for detection efficiencies, one may calculate the
occurrence rate of small planets, which reveal that the majority of
planets orbiting within 1 AU of solar-type stars, both those near (RV
surveys) and far (\ek survey), are smaller than Uranus and Neptune
(i.e., $<\,\sim$4 \rearthe), as described below.

\section{Occurrence Rates of 1--4 R$_{\oplus}$ Planets}

{\it Kepler} is superior to RV surveys for measuring occurrence
rates of planets down to 1 \rearth because it is better at 
detecting those planets. The Doppler reflex velocity of an 
Earth-size planet orbiting at 0.3 AU is only 0.2 \mse, 
difficult to detect with an observational precision of 1 \mse. 
But such Earth-size planets show up as a $\sim$10-sigma dimming
of the host star after co-adding the brightness measurements from each
transit.

The occurrence rate of Earth-size planets is a major goal of exoplanet
science. With three years of \ek photometry in hand, two groups worked to
account for the detection biases in \ek planet detection caused by
photometric noise, orbital inclination, and the completeness of the
\ek transiting-planet detection pipeline \cite{Howard2012,
  Fressin2013, Petigura2013a}. They found that within 0.25 AU of 
solar-type stars, small planets of 1--3x the size of Earth orbit
$\sim$30$\pm$5\% of Sun-like stars. In contrast, only 2$\pm$1\% have
larger planets of Neptune-size (4--6 \rearth), and only 0.5\% have
Jupiter-size planets (8--11 \rearth) orbiting that close
\cite{Howard2012, Petigura2013a}. Intriguingly, the occurrence rate
of close-in Jupiter-size planets found around stars in the \ek field
of view seems to be about half that found around nearby stars, a
difference not understood \cite{Wright2012}.

A new planet search of nearly 4 years of \ek photometry revealed
planets as small as 1 \rearth and orbital periods up to 200 days
\cite{Petigura2013b}. In this tour de force, they found 603 planets,
including 8 planets having sizes 1--2 \rearth that receive 1--4x the
incident stellar light flux that the Earth enjoys.  This new search
accounted for detectability efficiency of the smallest, Earth-size
planets by injecting into the \ek brightness measurements synthetic
dimmings caused by fake planets, and noting the detection success
rate.  This ``injection and recovery'' of fake Earth-size planets
yields a quantitative correction for efficiency, allowing
determination of the true occurrence rate of Earth-size planets.

Figure 1 shows the resulting fraction of Sun-like stars having planets
of different sizes \cite{Petigura2013b} with orbital periods of 5--100
days.   The lowest two bins show that 26.2\% of Sun-like stars have a
planets of size, 1--2 \rearthe, with orbital periods under 100 days.
Planets as large as Jupiter (11.2 \rearth) and Saturn (9.5 \rearth)
are more rare, occuring around less than $\sim$1\% of Sun-like stars in such orbits.
We do not know if the drop-off for the smallest planets is real, a
statistical fluctuation, or an incomplete bias correction

Figure 2 shows the resulting occurrence rate of planets around
Sun-like stars as a function of orbital period. The rate is about 15\%
at all orbital periods, within bins of multiples of orbital period
(i.e. 10--20d, 20--40d, 40--80d), as shown in Figure 2. This constant
planet occurrence with increasing orbital distance, in equal
logarithmic bins, surely informs planet formation theory.  Indeed, we
know of no theoretical cause of major discontinuities in planet
formation efficiency inside 1 AU.  No phase changes of major
planet-building material occur in that region.  A smoothly varying
occurrence rate, both observed and theoretically, supports mild
extrapolations of planet occurrence rates beyond orbital periods of
100 days where the measured rates are empirically secure \cite{Petigura2013b}.

Spectroscopy of the host stars of the Earth-size planets yields their
luminosities, providing a measure of the incident stellar light fluxes
falling on the planets.  This analysis shows 11\% of Sun-like stars
have a planet of 1--2 \rearth that receives 1-4x the incident stellar
flux that warms our Earth.  We note that all 10 such planets detected
in Petigura et al. orbit stars with sizes 0.5-0.8 solar radii,
i.e. smaller than the Sun.  The occurrence of Earth-size planet for
Sun-size stars may be somewhat different.  It is likely that a similar
number Sun-like stars (11\%) have 1--2 \rearth planets that receive
1/4--1x the incident flux that Earth enjoys.  {\em Thus, if one were
  to extrapolate to planets receiving 1/4--1x the incident flux of
  Earth, $\sim$22\% of Sun-like stars have a 1--2 \rearth planet that
  receives warming starlight within a factor of 4 of that enjoyed by
  our Earth, yielding similar surface temperatures, depending on
  surface reflectivity and greenhouse effects.}

\section{Properties:\ Masses, Radii, and Densities}

Though 1--4 \rearth planets are common, the theory of their interior
structures and chemical compositions is under active investigation
\cite{Seager2007, Fortney07, Zeng_Seager08, Rogers2011,
  Zeng_Sasselov2013, Lissauer2011, Lissauer2012, Fabrycky2012,
  Lopez2013b}. The measured radii, masses, and densities of small
planets constrain the relative amounts of iron and nickel, silicate
rock, water, and H and He gas inside the planets. Yet the measurements
of planet radius and mass leave degeneracies in the interior
composition. Even Uranus and Neptune, which have precisely measured
gravitational fields, have compositional degeneracies
\cite{Helled2013}. The interior compositions of small exoplanets are
similarly compromised by the possible different admixtures of the
rock, water, and gas.  Nonetheless, systematic correlations surely
exist between planet mass, radius, orbital distance, and stellar type
\cite{Fortney2013, Lopez2013b, Weiss2013, Weiss2014}, making
measurements of exoplanet radii and masses useful for understanding
the key processes of their formation.

Radii of exoplanets are measured based on the fractional dimming of
host stars as planets transit and are known for all \ek objects of
interest.  Planet masses require additional observations, and stem from
Doppler-measured reflex motion of the host star or from
variations in the time the planet crossing in front of the star each orbit
(transit-timing variations, TTV) caused by planets pulling
gravitationally on each other.

To date, 33 planets of 1--4 \rearth have measured radii and masses
with better than 2-$\sigma$ quoted accuracy.  The \ek Team recently
announced the masses and radii of 16 small transiting planets,
doubling the number of such well-studied planets \cite{Marcy2014}, and
the transit-timing variations of Kepler-11 planet system and other
Kepler Objects of Interest (KOI) have provided additional measured
masses \cite{Lissauer2013, Hadden2013, Weiss2014}.

Figure 3 shows two representative applications of the Doppler
technique to determine planet masses for Kepler-78 and Kepler-406.
Each star reveals repeated dimmings in \ek photometry due to
their transiting planets with orbital periods of 8.5 hours and 2.43 days
\cite{Batalha2013}, giving planet radii of 1.20 and 1.41 \rearthe,
respectively. Doppler measurements exhibit periodicites in phase with
the the orbit, yielding the reflex velocities of the star and hence the
masses of both planets, 1.69 and 4.71 \mearthe, respectively.
The resulting densities of the two planets are 5.3 $\pm$1.8 \gcc and
9.2$\pm$3.3 \gcc, respectively, both consistent with a purely rocky
interior \cite{Pepe2013, Howard2013b, Marcy2014}.  (For reference, the
Earth's bulk density is 5.5 \gcc.) These Doppler measurements are
expensive, requiring $\sim$45 minute exposures with the world's
largest telescopes on 50--100 different nights, while maintaining a
Doppler zero-point with a precision of 1 \mse, i.e., measuring
wavelengths to 9 significant digits.

In the analysis that follows, we include both Doppler-determined and
TTV-determined planet masses. It is worth noting that the TTV planet masses are
mostly lower than the RV-determined masses for given radii 
(though Doppler and TTV measurements of the same planets agree), 
for reasons not understood \cite{Weiss2014}.
Perhaps multi-planet systems, which allow TTV measurements, survive dynamically 
only if the planet masses are low enough to limit
catastrophic dynamical chaos. 

All 33 transiting 1--4 \rearth planets
with measured radii, ($>$2-$\sigma$) masses, and densities were vetted
in detail \cite{Weiss2014}. We explore the interdependencies among
these three measured quantities for
1--4 \rearth planets in Figures 4 and 5. 
Figure 4 shows planet density as a function 
of radius for all 33 known exoplanets smaller than 4 \rearth (dots).  
We also include Mars, Venus, and Earth (diamonds at left) and 
Uranus and Neptune (diamonds at right) as touchstones.
The planets reveal a dichotomy in their densities:\ those larger
than 2 \rearth are, with one exception, lower density than Earth, 
indicating their interiors contain substantial volumes of non-rocky, 
low-density material. 
For planets larger than 1.5 \rearthe, density declines with increasing radius; 
bigger planets have increasing amounts of low-density gas. 

By contrast, the smallest planets (1--1.5 \rearthe) all have measured 
densities above 5 \gcc, consistent with interiors of rock (silicate)
and iron-nickel. Indeed, though the scatter is large, the planets 
smaller than 1.5 \rearth have measured densities that increase with 
increasing radius (left side of Figure 4). The highest densities 
occur near a planet radius of $\sim$1.5 \rearth, at which value the 
average planet density is 7.6 \gcc \cite{Weiss2014, Rogers2014}, 
indicating purely rocky interiors.

Among the prominent examples of planets with size 1.5--4.0 \rearth and
sub-rocky densities are GJ 1214 b \cite{Maness07, Gillon2007,
  Torres2008, Charbonneau2009} with a radius of 2.68 \rearth and a
mass of only 6.55 \mearthe, yielding a bulk density of 1.87 \gcc.  For
comparison, Uranus and Neptune have densities of 1.27 and 1.63 \gcc,
well below Earth's (5.51 \gcc).  Similarly, the five inner planets
around Kepler-11, as well as the exoplanets GJ 3470 b, 55 Cnc e, and
Kepler-68 b all have densities less than 5 \gcc, with some under 1
\gcc \cite{Lissauer2013, Bonfils2012, Endl2012, Demory2013,
  Demory2011, Gilliland2013}. Thus, as shown in Figure 4, planets of
2--4 \rearth have densities too low to be mostly rock by volume.

Even larger planets, 4-6 \rearthe, have densities
that are even lower, near 1.0 \gcc \cite{Cochran2011,
  Hartman2011}. Jupiter and Saturn in our Solar System similarly have
densities near unity, due to large amounts of gas.  Similarly, the
sub-Earth bulk densities of planets larger than 2 \rearth indicate
that they contain significant amounts of H, He, and probably some
water \cite{Figueira2009, Rogers_Seager2010a, Batygin2013, Lopez2013}.

In contrast, the following planets with radii less 2 \rearth all have
2-$\sigma$ measured densities over 5 \gcc: CoRoT 7b, Kepler-10b,
Kepler-36b, KOI-1843.03, Kepler-78b, Kepler 406b, Kepler 100b, Kepler
113b, and Kepler 99b \cite{Queloz2009, Batalha2011, Carter2012,
  Rappaport2013,Sanchis-Ojeda2013, Pepe2013,Howard2013b, Marcy2014}.
These are the known rocky exoplanets, all validated as real at the
99\% confidence level.  All of them are smaller than 1.5 \rearthe.

Thus, we find a density dichotomy, with the dividing radius being near
1.5 \rearthe.  Planets smaller than 1.5 \rearthe have densities
consistent with a predominantly rocky interior, while those larger
than 1.5 \rearth appear to contain increasing amounts of gas with
increasing radius \cite{Weiss2013, Rogers2014, Lopez2013, Fortney2013,
  Lopez2013b}.

\section{Structure:\ Core-Envelope Model of 1--4 R$_{\oplus}$ Planets}

The two domains of 1--4 \rearthe planets, separated at 1.5 \rearthe,
motivate separate treatment of the mass-radius relationship in each
domain.  An empirical fit to the density-radius relation provides a
way to explore the ratio of rocky to low-density material in some
detail.  We fit a linear relation to density as function of radius for
all planets smaller than 1.5 \rearthe.  We restrict ourselves to a
linear relation in this domain because the density measurements have
large errors and because of the modest compressibility of rock.

In performing the weighted fit, we include all 22 exoplanets with
radius and mass measurements, regardless of the quality of the mass
measurement, to mitigate any bias in mass \cite{Weiss2014}.  This
linear fit includes the four solar system rocky planets with
uncertainties of 10\% in density so that they do not dominate the fit.
We note that both the exoplanets and solar system planets exhibit
an increase in density with increasing radius.  The mass-density
dependence for exoplanets is
anchored with Kepler-78b having $R$=1.2 \rearth and $\rho$=5.3--5.6
\gcc while the other exoplanets between 1.4--1.5 \rearth have mostly higher
densities between 7-14 \gcc, albeit with large uncertainties (Figure 4).

By including exoplanets having measured masses that are marginally significant, we
promote a statistically useful representation of planets of all masses
at a given planet radius \cite{Marcy2014, Weiss2014, Rogers2014}.
For all planets smaller than 1.5 \rearthe, a linear fit to density 
as a function of radius yields

$$\rho = 2.32 + 3.19 R/R_{\oplus} \ \ \ \ \ \ \ \  ({\rm for \ } R < 1.5 R_{\oplus})$$

\noindent as described in \cite{Weiss2014}. This linear relation is 
displayed as the dashed line in Figure 4, and is translated into a 
mass-radius relation in Figure 5.  The 
linear relation reveals a modest increase in density with increasing 
planet size up to 1.5 \rearthe, likely due to gravitational 
compression.  Among the exoplanets alone (without the solar system
planets) the apparent rise in density with radius hinges precariously on
the smallest exoplanet, Kepler-78b. We emphasize that the two constants in this linear
relation are heavily influenced by the terrestrial planets in our
Solar System that reside at larger orbital distances.  This linear
relation thus stems from a mélange of small planets orbiting both
close-in and farther out. 

For all planets larger than 1.5 \rearthe, a power-law fit to mass as a 
function of radius is adequate to accommodate the apparent curvature 
in the mass-radius measurements. The resulting power-law fit yields

$$M/M_{\oplus} = 2.69 (R/R_{\oplus})^{0.93} \ \ \ \ \ \ \ \  ({\rm for \ } R > 1.5 R_{\oplus})$$

\noindent as described in \cite{Weiss2014}. This mass-radius relation for
1.5--4.0 \rearth planets is shown as the solid line in the right half
of Figure 4. Planet density apparently declines with radius, indicating increasing 
amounts of low-density material as planet radius increases. The solid 
curve in the right half of Figure 4 ($R>$1.5\rearthe) resides 
systematically below the plotted points because the curve represents 
a power-law fit to {\em all} known exoplanets in that domain, while 
we have elected to plot only those points having mass measurements 
better than 2-$\sigma$ for visual clarity.

Figure 5 shows measured planet mass vs.\ radius for all 33 planets
having a mass measurement better than 2-$\sigma$. As in Figure 4, the
dashed line shows the previously described linear fit to density
vs.\ radius for all planets smaller than 1.5 \rearthe, likely composed
of mostly rocky material. We consider the existence of an envelope of 
low-density material on top of a rocky core by extending the dashed 
line to radii greater than 1.5 \rearthe. 

With such a linear extrapolation of the density relation, we can make 
an approximate prediction of the interior structure of planets larger 
than 1.5 \rearthe. At a given mass, the dashed line represents an
estimate of the size of the planet's rocky core. The size of a 
planet's low-density envelope, therefore, is represented by the horizontal 
distance between the dashed line and the plotted point for that planet. Consider the two examples 
of GJ 1214b and Kepler-94b, with dotted lines drawn from the planet's 
location in mass-radius space back to the radius representing their 
rocky cores (dashed line). The lengths of the dotted lines represent 
the additional radius, on top of any rocky core, that must consist of 
low density material to explain the enlarged radius at a given mass.

Thus, cloud of planets residing to the right of the ``rocky''
dashed line in Figure 5 support a model of exoplanet structure with
both rock and volatiles.   
These planets have larger radii (and volumes) than can be
explained by a purely rocky interior. Therefore, these planets 
surely contain large amounts of gas and ices to account for their 
large size, given their mass. {\em Clearly, the planets larger
  than 2 \rearth are composed of large contributions of gas in
  addition to any rocky core.} 

A core-envelope model follows from the expectation that the more dense
material will sink (differentiate) toward the center of the planet.
The argument presented here for large amounts of low density material
on a rocky core does not make use of any theoretical equation of
state. The low-density material, presumably H and He gas, must exist
in the planets larger than 2 \rearth on observational grounds alone.

\section{Interiors, Formation and Evolution}

The range of sizes of rocky planets is visible in Figures 4 and 5 as the observed rise
in density and mass with increasing radius for planets smaller than 1.5 \rearthe.
{\em It is an extraordinary accomplishment in planetary astrophysics that
  the accurately determined radii, masses, and densities of planets smaller than 
  1.5 \rearthe reveal increasing mass with radius, signalling their rocky
  interiors and associated gravitational compression.}  A linch-pin is
Kepler-78b that has radius 1.2 \rearth and density 5.3 \gcc,
compared to the handful of exoplanets of radius 1.4--1.5 \rearth that all
have higher densities, displaying gravitational compression and
supporting the linear relation for $\rho(R)$ in Figures 4 and 5.
Of course, Mars, Venus, and Earth also exhibit increasing density with
radius, offering further support.

For those planets larger than 1.5 \rearthe, the dramatically decreasing
density with increasing radius, visible in Figure 4, 
clearly indicates increasing amounts of volatiles. 
Extrapolating the mass-radius relation for purely rocky planets gives
an approximate division of the core and envelope for these ``mini-Neptunes.'' 
The dotted lines in Figure 5 give an example of this division. But that
division is certainly too simple:\ planets of a given radius must also 
have a diversity of rocky core masses and radii \cite{Lopez2013b, Weiss2014}.  
Because most of the mass resides in the core, not the gaseous envelope, 
only a diversity of rocky core sizes can explain, at a given radius, the
observed spread of planets masses. {\em Thus, the cloud of points in the
  right halves of Figures 4 and 5 represent planets with a range
  of both core masses and volatile content.}

The existence of two planet domains on either side of 1.5 \rearth are consistent
with planet formation models that suppose an accumulation of rocky
material up to some critical rocky core mass, followed by accretion of H and He gas. The sequence of
planets from 1--4 \rearth is then interpreted as a sequence of
various amounts of iron-nickel and rocky material with either none or
increasing amounts of accreted gas \cite{Chiang_Laughlin2013,
  Hansen_Murray2013, Lopez2012, Lopez2013, Gillon2012, Demory2012,
  Mordasini2012a, Lammer2013b, Raymond_Cossou2014}.

The spread in planet bulk densities at a given radius or mass may also
be due to the subsequent photo-evaporation of volatiles. Such
evaporation may be germane because nearly all of the 1--4 \rearth
planets described here orbit within 0.1 AU of a host FGKM-type star,
and therefore their envelopes would be subject to heating, UV
deposition, and atmospheric escape \cite{Demory2012, Gillon2012,
  Lammer2013a, Lopez2013}.
These mechanisms for loss of envelopes, along with models of in situ formation of mini-Neptunes, seem to predict the range of sizes, masses, and densities that are 
observed for the 1--4 \rearth planets \cite{Lopez2013b}. Detailed models of planet 
interiors, including the range of chemical compositions, stratified 
differentiation, and equations of state are needed to predict the 
plausible bulk densities associated with planets with a given mass 
\cite{Rogers_Seager2010b, Rogers2011, Zeng_Sasselov2013, Lopez2013b, 
  Rogers2014}.

\section{Correlations with Heavy Element Abundance}
The abundances of heavy elements in the protoplanetary disks around
young stars may influence the efficiency of formation of the rocky cores made of
such elements.    Spectra of the brightest \ek host stars of transiting planets were
analyzed by \cite{Buchhave2013} to yield their abundances of heavy
elements relative to the Sun (``metallicities''). The planets with sizes 
greater than 3.5 \rearth orbit host stars that have, on average, high 
metallicity:\ they are rich in heavy elements relative to the Sun.
Figure 6 shows the metallicity on a log scale (zero being solar) of 
over 400 stars that host $\sim$600 \ek exoplanets \cite{Buchhave2013}.. Figure 6 shows 
that planetary systems seem to fall into three populations defined by
different radii and associated stellar metallicities. The smallest planets
($R<$1.7 \rearth) have, on average, host stars with metallicities slightly
less than that of the Sun. The largest planets
($R>$3.4 \rearth) orbit stars having systematically higher
metallicities than the Sun.

One possible explanation for this correlation between planet size and
the metallicity of the host star is that giant planets are created
from a rocky core that accretes H and He gas from the protoplanetary
disk.  But the gas in protoplanetary disks disipates quickly (within a
few million years).  The heavy elements in the protoplanetary disk must
form a rocky core quickly enough to accrete the gas before it
vanishes.  If so, the core can accrete H and He gas to form the low
density, gaseous planet.  Those stars (and their protoplanetary disks)
that have only modest metallicity (or less) form rocky cores more
slowly, after most of the gas in the protoplanetary disk has vanished,
leaving only rocky cores that are devoid of a gaseous envelope
\cite{Buchhave2013}.  If this explanation is roughly correct, the
Earth resides at a planetary sweet spot, coming from a protoplanetary
disk with inadequate heavy elements to grow quickly enough to grab
huge amounts of gas, but adequate to initiate complex biochemistry.

\section{Habitable Zone:\ Humility and Hubris}

Scientific knowledge of complex systems is normally anchored by, and
repeatedly tested by, experimental evidence. The planetary conditions
necessary for biology certainly qualify as a complex physical,
chemical, and biological problem. A common construct toward such
discussions is the ``habitable zone,'' the orbital domain around a
star where life can arise and flourish. Unfortunately, we have no
empirical evidence of life arising, nor of it flourishing, around any
other star.

Such lack of experimental evidence of life has not slowed the debate
about the exact location of the habitable zone around stars of
different types. The passion exhibited in this debate is worth some
caution. We have no evidence of microbial life at any orbital
location within our solar system beside the Earth. We have no
empirical information about microbial life as a function of orbital
distance from our Sun or from any other star. We also have no evidence
of multicellular life around any other star, nor evidence of
intelligent life.

Thus, we have no empirical knowledge about the actual domain of
habitable zones, for any type of life, around any type of star. 
Moreover we have virtually no theoretical underpinnings about
exobiology. We still do not know how biology started on Earth. We do
not know the mechanisms that caused a transition from chemistry to
biology, nor do we know the biochemical steps that spawn proteins,
RNA, DNA, or cell membranes \cite{Lal2008}, though there is progress
\cite{Adamala2013}. Indeed, we still have a poor definition of life 
\cite{Szostak2012}. 

Our ignorance about both the necessary planetary environments and the
complex biochemical pathways for life should urge caution in
predicting, with multiple significant digits, the location of the
``habitable zones'' around other stars. We can't predict if Mars,
Europa, or Enceladus have habitable environments any better than we can
predict the weather in our home town a week in advance.

What is needed is a census of biology among a sample of nearby stars,
measuring the orbital locations and geological types of planets where
biologies exist. A door-to-door census of life among stellar
neighbors is needed to answer empirically and with credibility the
true domain of habitability around other stars. That census can be
carried out three ways:\ within our Solar System among water-bearing
planets and moons, by space-borne telescopes that perform chemical
assays of resolved rocky planets, and by searches for transmissions
from technological beings.

\section{Acknowlegments}
We thank Leslie Rogers, Eric Lopez, Jonathan Fortney, Dimitar
Sasselov, Jack Lissauer, Eugene Chiang, Greg Laughlin, and Sara Seager
for valuable conversations. We thank the extraordinary group of
engineers and scientists who worked tirelessly to produce the \ek
mission. \ek was competitively selected as the tenth NASA Discovery
mission. Funding for this mission is provided by the NASA Science
Mission Directorate. Some of the data presented herein were obtained
at the W.~M.\ Keck Observatory, which is operated as a scientific
partnership among the California Institute of Technology, the
University of California, and the National Aeronautics and Space
Administration. The Keck Observatory was made possible by the generous
financial support of the W.~M.\ Keck Foundation. We thank the many
observers who contributed to the measurements reported here. We thank
the NSF Graduate Research Fellowship, grant DGE 1106400.  This
research has made use of the NASA Exoplanet Archive, which is operated
by the California Institute of Technology, under contract with the
National Aeronautics and Space Administration under the Exoplanet
Exploration Program.  Finally, the authors wish to extend special
thanks to those of Hawai`ian ancestry on whose sacred mountain of
Mauna Kea we are privileged to be guests.  Without their generous
hospitality, the Keck observations presented herein would not have
been possible.


\bibliographystyle{pnas2009}            


\end{article}


\begin{figure}
\centerline{\includegraphics[width=0.7\textwidth]{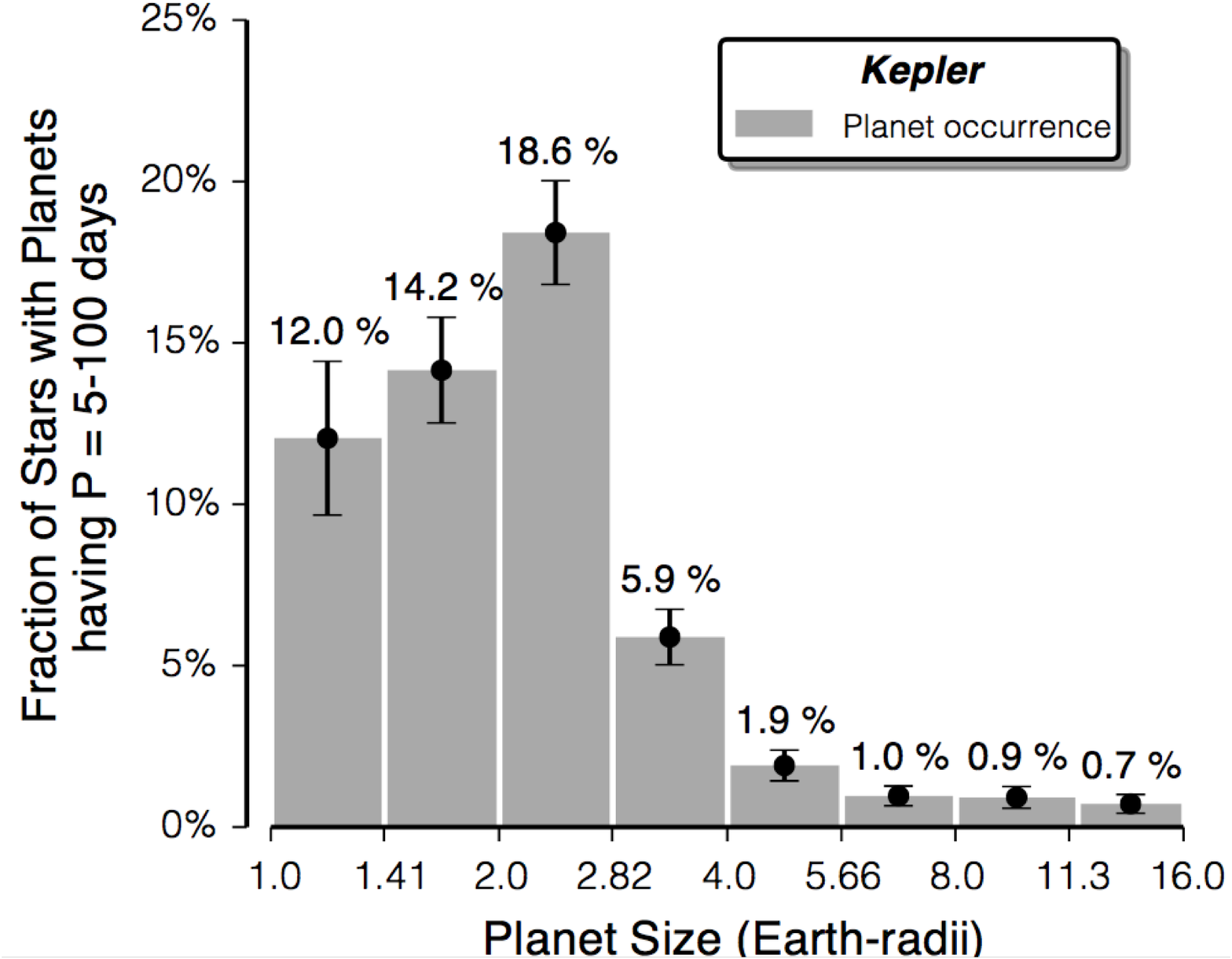}}
\caption{The size distribution for planets around Sun-like stars.
  The fraction of Sun-like stars (G- and K-type) hosting planets of
  a given planet radius are tallied in equal logarithmic bins. Only 
  planets with orbital periods of 5--100 days (corresponding to orbital 
  distances of 0.05--0.42 AU) are included. Together, the lowest two 
  bins show that 26\% of Sun-like stars have planets of 1--2 \rearth 
  orbiting within $\sim$0.4 AU. The occurrences of Neptune-size planets
  (2.8--4 \rearth) and gas-giant planets (8--11 \rearth) are 5.9\% and
  0.9\%, respectively, more rare than Earth-size planets
  \cite{Petigura2013b}.}
\label{fig:petigura_size_dist}
\end{figure}

\begin{figure}
\centerline{\includegraphics[width=0.7\textwidth]{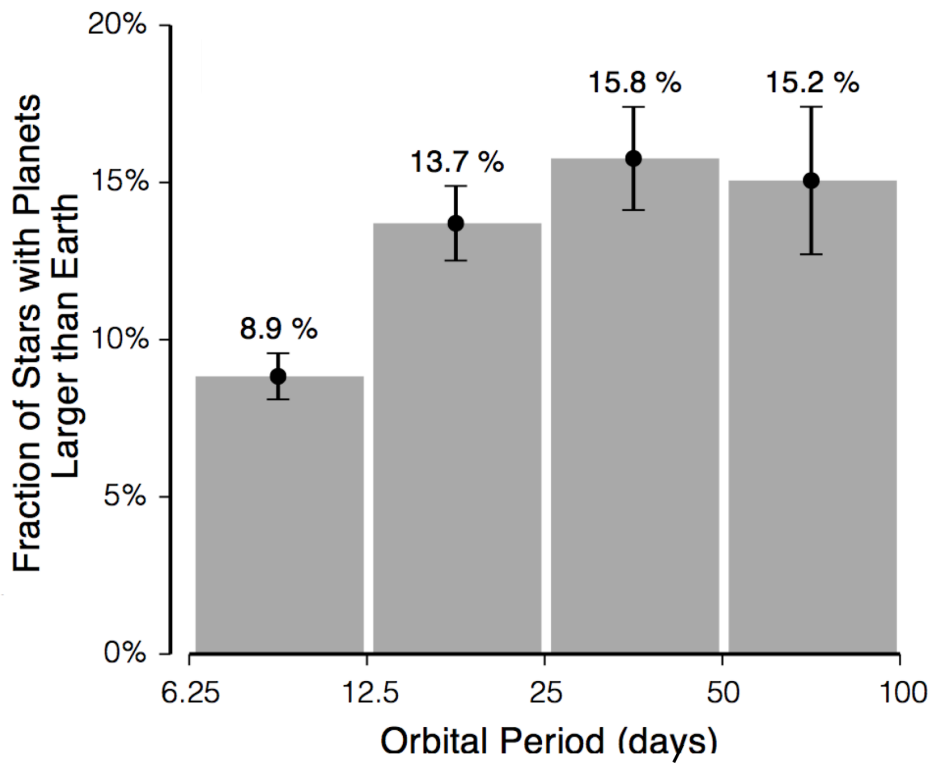}}
\caption{The fraction of Sun-like stars having planets larger than
  Earth and within $\sim$0.4 AU, 
  as a function of the planets' orbital periods (log scale). 
  The occurrence of planets is roughly constant, $\sim$15\%, in period 
  bins sized by equal factors of 2 in orbital period between 12--100 days. 
  Thus, planet occurrence is roughly constant with orbital distance, 
  d$N$/d$\log a$=constant, in the inner regions of planetary systems
  \cite{Petigura2013b}.}
\label{fig:petigura_period_dist}
\end{figure}

\begin{figure}
\centerline{\includegraphics[width=0.35\textwidth]{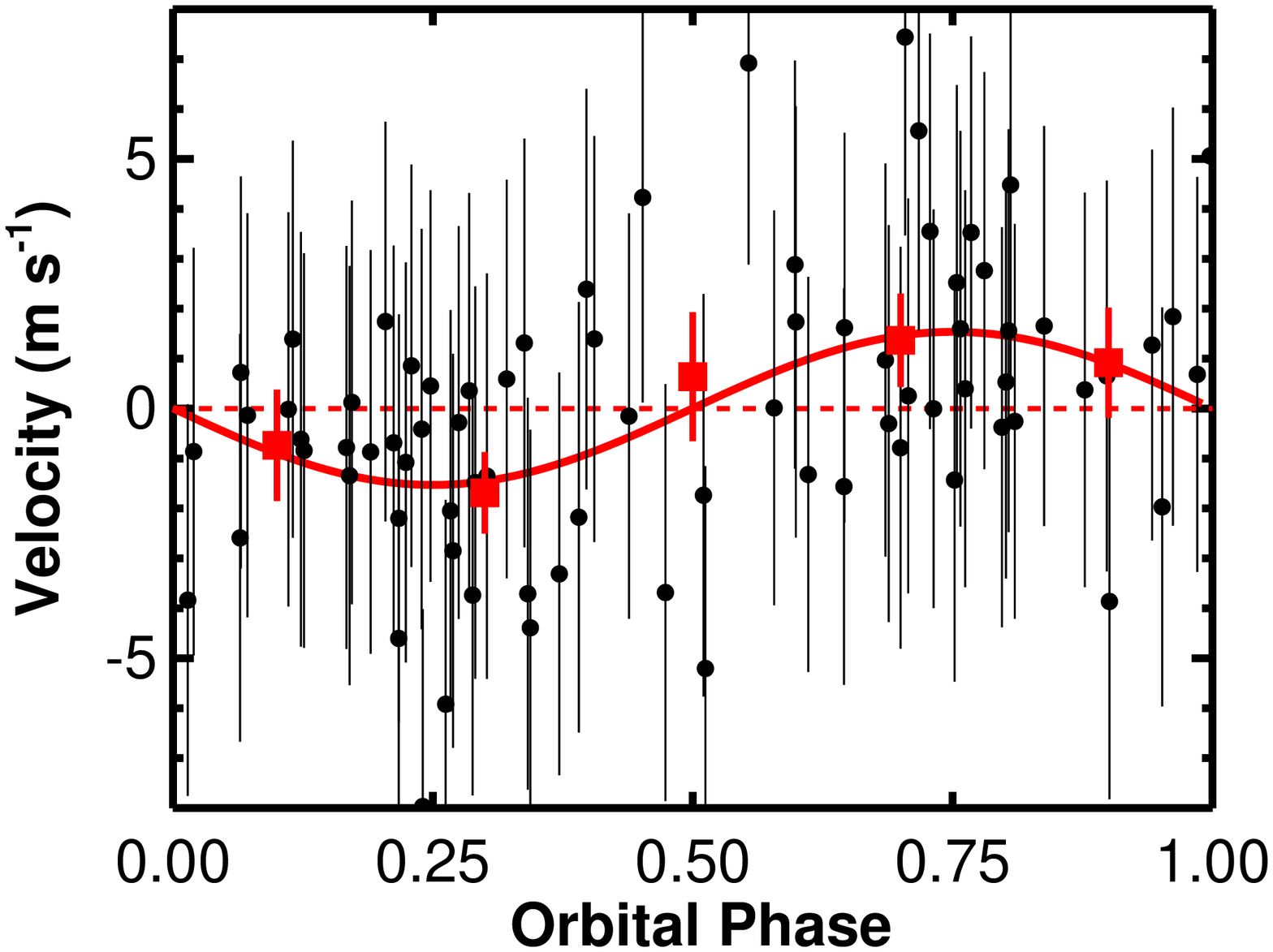}\includegraphics[width=0.35\textwidth]{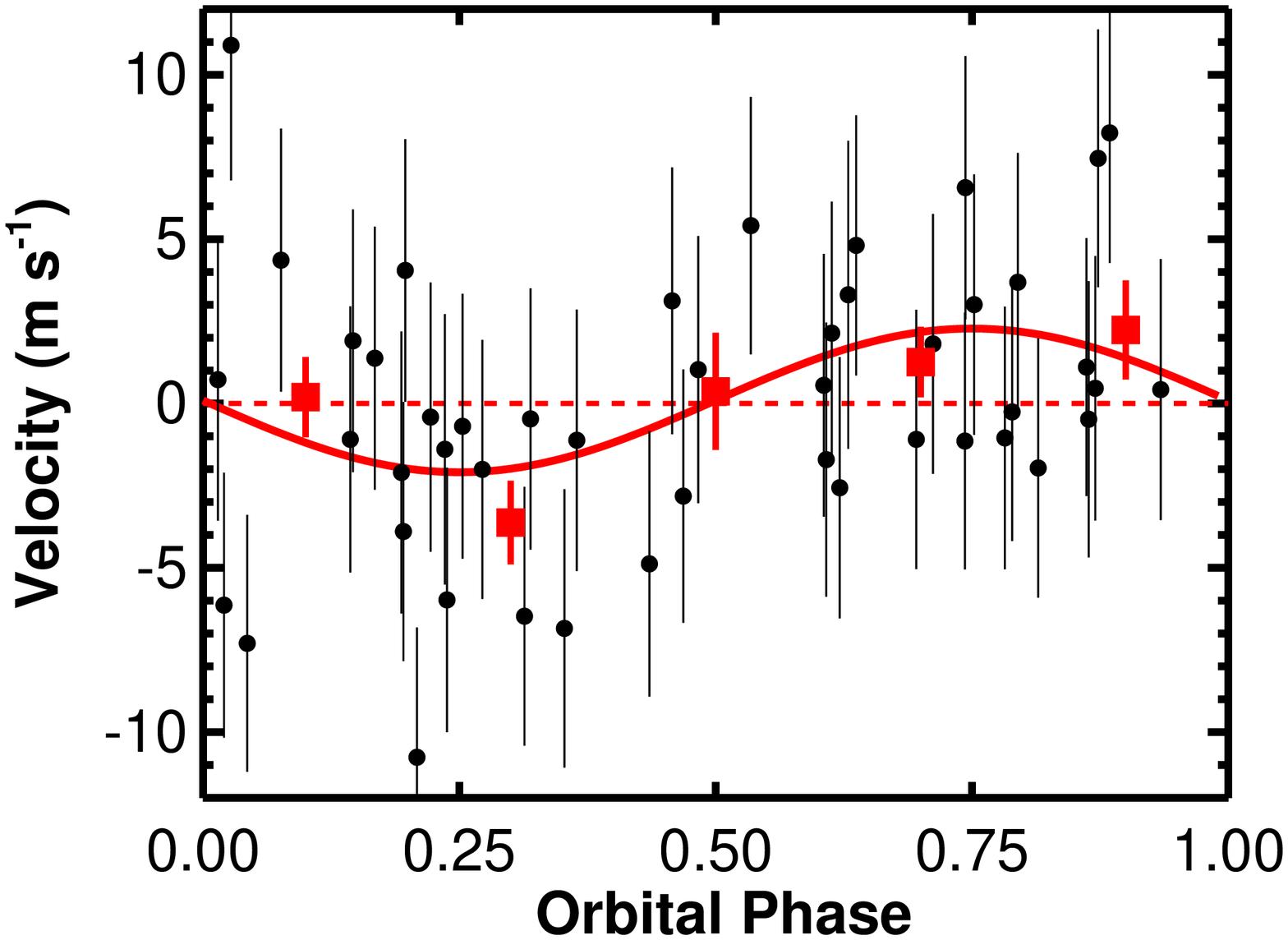}}
\caption{Doppler measurements made during the orbits of the exoplanets
  Kepler-78 (left) and Kepler-406 (right), stars that harbor planets with
  radii of 1.20 and 1.41 \rearthe, respectively. The Doppler
  measurements show a sinusoidal periodicity, yielding masses corresponding 
  to densities of 5.3$\pm$1.8 \gcc and 9.2 $\pm$3.3 \gcc, implying rocky 
  compositions \cite{Howard2013b, Marcy2014}.}
\label{Doppler}
\end{figure}

\begin{figure}
\centerline{\includegraphics[width=0.7\textwidth]{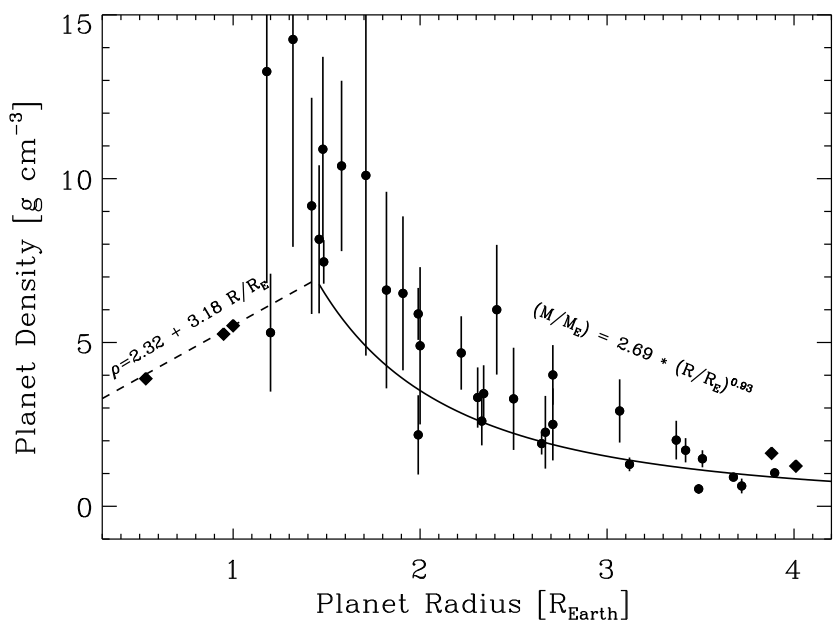}}
\caption{Planet density vs.\ radius for all 33 known exoplanets smaller
  than 4 \rearthe that have 2-$\sigma$ mass determinations. Venus,
  Earth, Mars, Uranus, and Neptune are included (diamonds). The radius
  of $\sim$1.5 \rearth has the highest densities, and marks the transition between rocky
  planets (smaller size, at left) and planets with increasing amounts
  of low density material (larger size, at right) \cite{Lopez2013b,
    Weiss2014, Rogers2014}. For radii 0--1.5 \rearthe, density
  increases with planet radius, consistent with a purely rocky
  constitution. In the radius range of 1.5--4.0 \rearthe, density 
  decreases with radius, indicating increasing amounts of H and He gas 
  or water. The transition radius at 1.5 \rearth has a density maximum 
  near $\sim$7.6 \gcc (weighted average). A linear fit including all
  planets (including sub-2-$\sigma$ densities, not shown) for  
  $R<1.5$ \rearth (dashed line) yields: 
  $\rho(R) = 2.32 + 3.19 R/R_{\oplus}$ 
  in units of \gcc. A fit for $R>1.5$ \rearth
  (solid line) yields a density law: 
  $\rho(R) = 2.69 (R/R_{\oplus})^{0.93}$ 
  in \gcc, consistent with a characteristic
  core mass of roughly 10 \mearth \cite{Lopez2013b, Weiss2014}.}
\label{fig:rho_vs_rad_ss}
\end{figure}

\begin{figure}
\centerline{\includegraphics[width=0.7\textwidth]{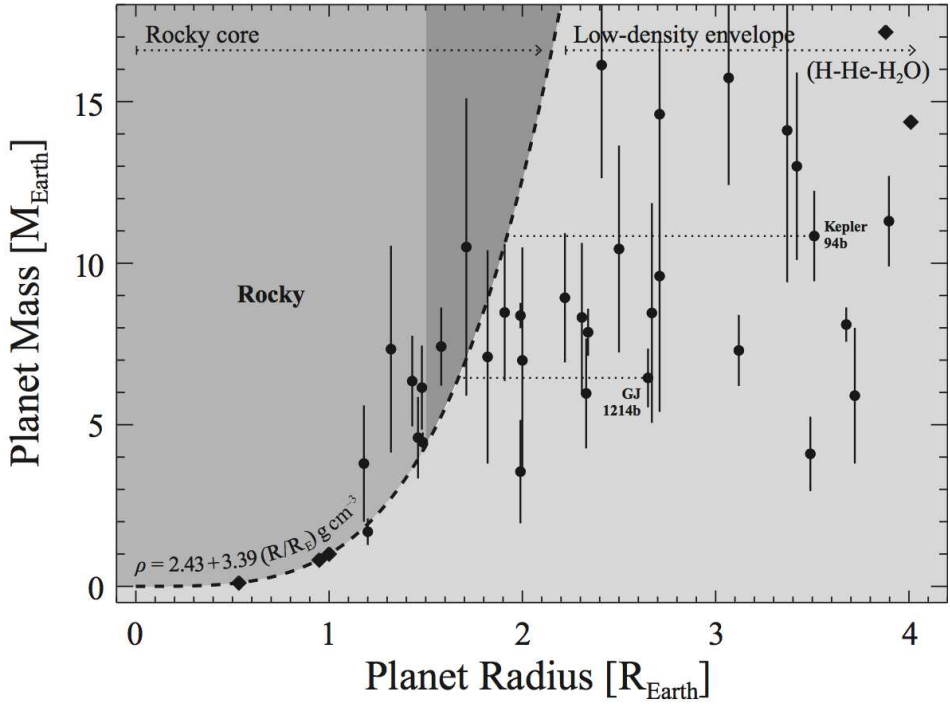}}
\caption{Planet mass vs.\ radius, including both the 33 known
  exoplanets smaller than 4 \rearthe with 2-$\sigma$ mass
  determinations (circles) and the solar system planets (diamonds).  
  Planet mass is correlated with radius in the domain $R<1.5$ \rearthe.  
  The dashed line marked ``rocky'' represents the linear density-radius
  relation from Figure 4, projected into mass-radius space. 
  The points residing near that dashed line represent planets that
  must be mostly rocky. The points residing to the right of the
  ``rocky'' dashed line represent planets with radii too large to be
  purely rocky. For such planets, dashed line represents a simple 
  approximation of the  
  dividing line between a rocky core and a low-density envelope:\ the 
  horizontal distance to the left of the dashed line (dark gray) represents
  the radius of the rocky core, while the horizontal distance to the 
  left of the dashed line (light gray) represents the extra radius 
  from the low density material (H and He or water) in 
  the envelope, which contributes extra size but
  negligible mass; see \cite{Lopez2013b, Weiss2014, Rogers2014,
    Fortney2013, Demory2012, Gillon2012, Lammer2013a}. As an example, the 
  additional size, on top of the rocky core, contributed by the H and He 
  or H$_2$O envelopes for GJ 1214b and for Kepler-94b are indicated by 
  dotted lines. Planets of 1--4 \rearth are well modeled by a rocky
  core containing most of the mass plus a low-density envelope, if
  any, that enlarges the planet's radius.}
\label{fig:mass_detected_vs_radius_ss}
\end{figure}

\begin{figure}
\centerline{\includegraphics[width=0.6\textwidth]{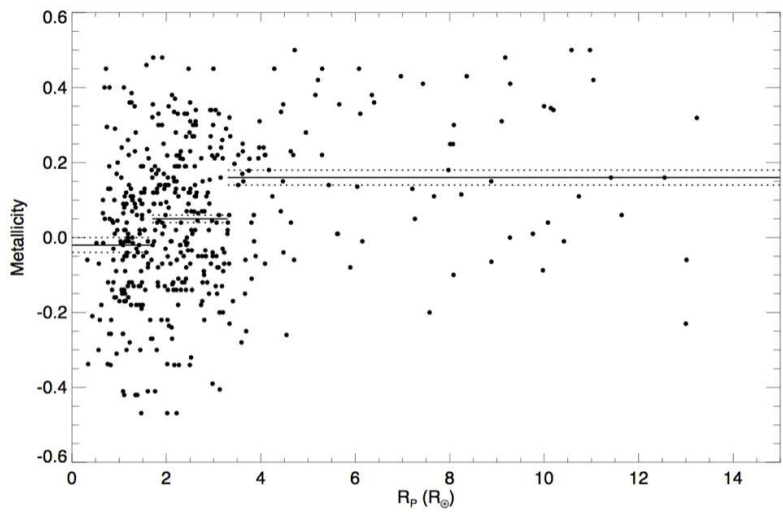}}
\caption{Abundance of heavy elements (metallicity) of the host star
  vs.\ planet radius for over 400 stars as a function of the size of
  the \ek planet orbiting it. The planets with sizes larger than
  4 \rearth have host stars relatively rich in heavy elements. In
  contrast the smaller planets orbit stars that are roughly solar-like in
  metallicity.  The explanation may be that high metalicity in the
  proplanetary disk allowed rocky cores to form quickly, before the
  gas in the disk vanished, allowing the cores to gravitationally
  accrete that gas to make gas-rich planets.}
\label{Buchhave}
\end{figure}






\end{document}